\documentclass[longnamesfirst,apjl]{emulateapj}
\usepackage{apjfonts,natbib,epsfig}
\tighten

\newcommand{\beq}{\begin{equation}}
\newcommand{\eeq}{\end{equation}}
\newcommand{\beqa}{\begin{eqnarray}}
\newcommand{\eeqa}{\end{eqnarray}}
\newcommand{\fsky}{f_{\rm sky}}
\begin{document}

%\title{Cosmological Applications of Weakly Lensed Supernovae}
\title{Cosmology from supernova magnification maps}

\author{Asantha Cooray$^1$, Daniel E. Holz$^{2,3}$, and Dragan Huterer$^3$}
\affil{$^1$Department of Physics and Astronomy, University of California, Irvine, CA 92617\\
$^2$Theoretical Division, Los Alamos National Laboratory, Los Alamos, NM 87545\\
$^3$Kavli Institute for Cosmological Physics and
  Department of Astronomy and Astrophysics, University of Chicago, Chicago, IL 60637}

%\righthead{Weak Lensing of Supernovae}
\righthead{Supernova magnification maps}
\lefthead{COORAY ET AL.}
\begin{abstract}
High-z Type Ia supernovae are expected to be gravitationally lensed by the foreground
distribution of large-scale structure. The resulting magnification of
supernovae is statistically measurable, and the angular correlation of the
magnification pattern directly probes the integrated mass density along the
line of sight. Measurements of cosmic magnification of supernovae therefore complements
galaxy shear measurements in
providing a direct measure of clustering of the dark matter.
As the number of supernovae is typically much smaller than
the number of sheared galaxies, the
two-point correlation function of lensed Type Ia supernovae suffers from
significantly increased shot noise. 
%This is only partially compensated for by the smaller
%errors on each individual supernova measurement, as compared to the intrinsic shape error
%in individual galaxy shear measurements.
Neverthless, we find that the magnification map of a large sample of
supernovae, such as that expected from next generation dedicated searches, will
be easily measurable and provide an important cosmological tool.  For example,
a search over 20 sq.\ deg.\ over five years leading to a sample of $\sim$10,000
supernovae would measure the angular power spectrum of cosmic magnification
with a cumulative signal-to-noise ratio of $\sim$20. This detection can be further
improved once the supernova distance measurements are
cross-correlated with measurements of the foreground galaxy distribution.  The
magnification maps made using supernovae can be used for important cross-checks
with traditional lensing shear statistics obtained in the same fields, as well as
help to control systematics. We discuss two applications of supernova magnification maps:
the breaking of the mass-sheet degeneracy when estimating masses of shear-detected
clusters, and constraining the second-order corrections to weak lensing
observables.

\end{abstract}

\section{Introduction}
\label{sec:introduction}
Type Ia supernovae (SNe) are by now firmly established as powerful probes of
the expansion history of the universe \citep{Baretal04,Knop:03,Riess:04}. In
particular, the luminosity distance measurements from SNe provide a direct
probe of dark energy in the universe and its temporal behavior (e.g.,
\citealt{Huterer:05} and references therein). Numerous current and future SNa
Ia surveys are being planned or performed, and this community-wide effort is
expected to reach its apex with the NASA/DOE Joint Dark Energy Mission (JDEM).

While SNe are very good ($\sim10\%$ errors in flux) standard candles, the
inferred luminosity of a given supernova is affected by cosmic magnification
due to gravitational lensing from the mass distribution of the large scale
structure along the line of sight between the supernova and the observer
\citep{Frieman:97,HolzWald:98}.
%Thus, a fundamental limitation in measuring luminosity distance to a given
%supernova is the fluctuation of the SN flux caused by slight bending of light
%of the SN by the intervening large-scale structure.
This is a fundamental limitation to the utility of standard candles, and SNe at high
redshift ($z>1$) are especially prone to fluctuations of their flux due to
lensing. There was a recent claim of evidence for weak lensing of SNe
from the \citet{Riess:04} sample \citep{Wang:05}; however, this claim remains unconfirmed
as the correlation with
foreground galaxies that would be expected from lensing has not been observed
\citep{Menard}.  

Weak lensing biases the luminosity measurement from each SN and thus introduces
a systematic error in the extraction of cosmological parameters.  With large
number of SNe in each redshift interval at high $z$ this systematic can be
essentially averaged out \citep{Dalal:03,Holz:04}, though the full lensing
covariance must be taken into account for accurate cosmological parameter
estimates (Cooray et al. 2005).  While lensing has mostly been considered as a
nuisance, planned large area SN surveys provide an opportunity to treat lensing
magnification on SNe as a signal, and extract information about the underlying
dark matter distribution. In practice, by comparing the SN Hubble diagram
averaged over all directions with individual SN luminosity distance
measurements, one can map out anisotropy in the SN Hubble diagram.  This
anisotropy will trace cosmic magnification, or in the weak gravitational
lensing limit, it will be linearly proportional to the convergence and hence to
fluctuations in the distribution of the foreground large scale structure.

Previous studies have considered potential applications of cosmic convergence
as a probe of the large-scale dark matter distribution \citep{Jain:02}. Cosmic 
magnification has already been detected via cross-correlation
between fluctuations in background source counts, such as quasars or X-ray
sources, and a sample of low-redshift foreground galaxies (see
\citet{Bartelmann:01} for a review).  Past detections
have often been affected by systematic uncertainties, but the powerful
Sloan Digital Sky Survey (SDSS) has recently made the first reliable detection of
cosmic magnification \citep{Scranton:05}.  Nevertheless, even this measurement
using several hundred thousand quasars and upwards of ten million
foreground galaxies was at a relatively modest 8$\sigma$ level, illustrating
the intrinsic difficulty in extracting the cosmic magnification. Some proposals
for future detections involve the use of 21 cm background anisotropies of the
hydrogen distribution prior to reionization \citep{Zhang:05}, but these studies
are experimentally challenging and affected by large theoretical uncertainties
in the amplitude of the expected signal and its modification due to lensing
\citep{Cooray:04}.

In this {\it Letter} we propose mapping cosmic magnification with a sample
of Type Ia SNe.  We compute the predicted angular power spectrum of
lensing magnification and estimate how accurately it can be measured, as well
as how this measurement can be improved through cross-correlating supernova distances
with the foreground distribution of galaxies. We envision several important applications of
this technique.

The {\it Letter} is organized as follows: In \S~\ref{sec:lensing} we
describe the weak lensing of SNe and the extent to which it can be
measured from future SN searches. In \S~\ref{sec:discussion} we
discuss our results and comment on specific applications of SN magnification maps. We
adopt the current concordance cosmological model with a
Hubble constant of $h=0.7$, matter density $\Omega_m=0.3$, cosmological
constant $\Lambda=0.7$, and normalization of the matter power spectrum 
$\sigma_8=0.85$.

\begin{figure}[!t]
\psfig{file=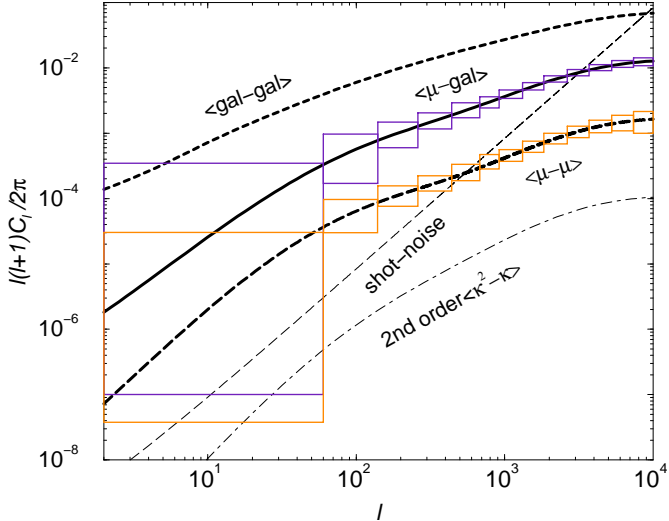,width=2.8in, angle=-90}
\caption{Angular power spectrum of cosmic magnification (bottom long-dashed curve),
cross-correlation between magnification and foreground galaxies (middle solid
curve), and foreground galaxy clustering (top short-dashed curve).  Fractional error
in the luminosity of each SN has been assumed to be 0.1, and the error boxes
account for both the sample (cosmic) variance due to the limited survey area
and the presence of shot noise due to the finite number of SNe. We assume
10,000 SNe obtained over an area of 10 sq.\ deg.\ with a uniform
distribution in redshift between 0.1 and 1.7. The thin dashed line shows the
shot noise in the magnification power spectrum.  The cross-correlation makes
use of the sample of SNe at $z>0.7$, while the foreground galaxy sample
is from \citet{Scranton:05}. The thin dot-dashed line shows the second-order
correction to magnification following \citet{Menard:03} and using the halo
model to describe the density field bispectrum.}
\label{fig:cl}
\end{figure}

\section{Weak Lensing of Supernovae}
\label{sec:lensing}

We begin by summarizing the effect of lensing magnification on SNe.  While we
concentrate on SNe, which have been firmly established as important and
reliable cosmological probes, our study applies to any standard
candle~(e.g.~\citet{HolzHughes05}).  Luminosity of a given supernova at a
redshift $z$ and located in the direction $\hat{\bf n}$, $L(z, \hat{\bf n})$,
is affected by weak lensing magnification so that $L(z, \hat{\bf n}) = \mu(z,
\hat{\bf n}) \bar{L}$, where $\mu(z, \hat{\bf n})$ is the weak lensing induced
magnification in the direction $\hat{\bf n}$ and at redshift $z$, and $\bar{L}$
is the true luminosity of the supernova. Note that $\mu$ can take values
between the empty-beam value and infinity; the probability distribution
function of $\mu$, $P(\mu)$, has been extensively studied both analytically and
numerically (e.g., \citealt{Holz:98,Wang:02}). Since $\langle \mu \rangle =1$,
one can average over large samples to determine the mean luminosity $\bar{L}$
\citep{Wang:00,Holz:04}.  One can now consider spatial fluctuations in the
luminosity
\begin{equation}
\delta_L(z, \hat{\bf n}) = \frac{L(z, \hat{\bf n}) - \bar{L}}{\bar{L}} \, ,
\label{eq:delta_L}
\end{equation}
which traces fluctuations in the cosmic magnification $\mu$.  In the weak
lensing limit ($\mu, \kappa \ll 1$) we have
\begin{equation}
\mu = [(1-\kappa)^2-|\gamma|^2]^{-1} \approx 1 + 2 \kappa + 3\kappa^2+|\gamma|^2+...\, ,
\label{eq:higher_order}
\end{equation}
where $\kappa$ is the lensing convergence and
$|\gamma|=\sqrt{\gamma_1^2+\gamma_2^2}$ is the total shear. To first order in
the convergence, $\delta_L(z, \hat{\bf n})$ traces spatial fluctuations of
$2\kappa$, though higher order corrections may be important \citep{Menard:03}.
Traditional weak lensing involves measurement of statistics of the shear,
$\gamma_i$, as this leads to a distortion of background galaxy shapes
\citep{Bartelmann:01}.  Magnification, on the other hand, changes images sizes,
but suffers from the problem that the true size of cosmological objects is
highly uncertain.  Fluctuations in the luminosity of standard candles provides
a reliable way to probe cosmic magnification.

The angular power spectrum of magnification fluctuations is, assuming
statistical isotropy
\begin{equation}
\langle \mu_{\ell m}^\star \mu_{\ell' m'}\rangle = C_\ell^{\mu \mu} \delta_{\ell \ell'} \delta_{m m'} \, .
\end{equation}
where $\mu_{\ell m}$ are the multipole moments of the magnification. Using
the Limber approximation, the angular power spectrum can be written as \citep{Kaiser:98,Cooray:00}
\begin{eqnarray}
C_\ell^{\mu \mu} &=& \int dr  \frac{W^2(r)}{d_A^2} P_{\rm dm}\left(k=\frac{\ell}{d_A},r\right) 
\nonumber \\
W(r) &=& 3 \int dr' n(r') \Omega_m \frac{H_0^2}{c^2 a(r)} \frac{d_A(r) d_A(r'-r)}{d_A(r')}
\label{eq:cl}
\end{eqnarray}
where $r$ is the comoving distance, $d_A$ is the angular diameter distance and
$n(r)$ is the radial distribution of SNe normalized so that $\int dr\,
n(r) = 1$. $P_{\rm dm}(k, r)$ is the three-dimensional power spectrum of dark
matter evaluated at the distance $r$; we calculate it using the halo model
of the large-scale structure mass distribution \citep{Cooray:02}.
The next order correction term, $\langle \kappa^2_{\ell m}
\kappa_{\ell' m'}\rangle$, is easily related to the convergence bispectrum
and we calculate it using the halo model \citep{Cooray:00}.

\begin{figure*}[!t]
\centerline{
\psfig{file=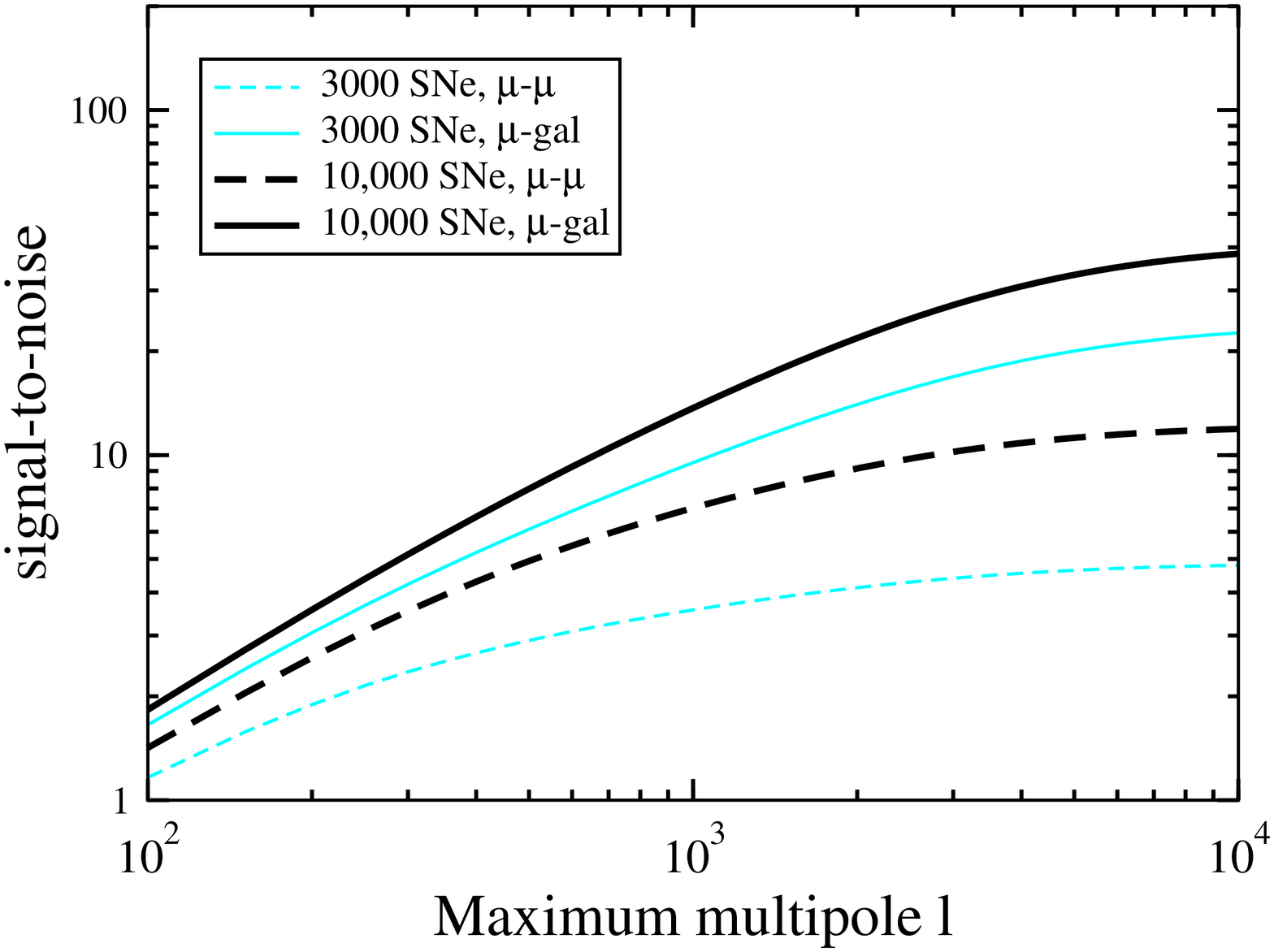,width=2.8in, angle=-90}
\psfig{file=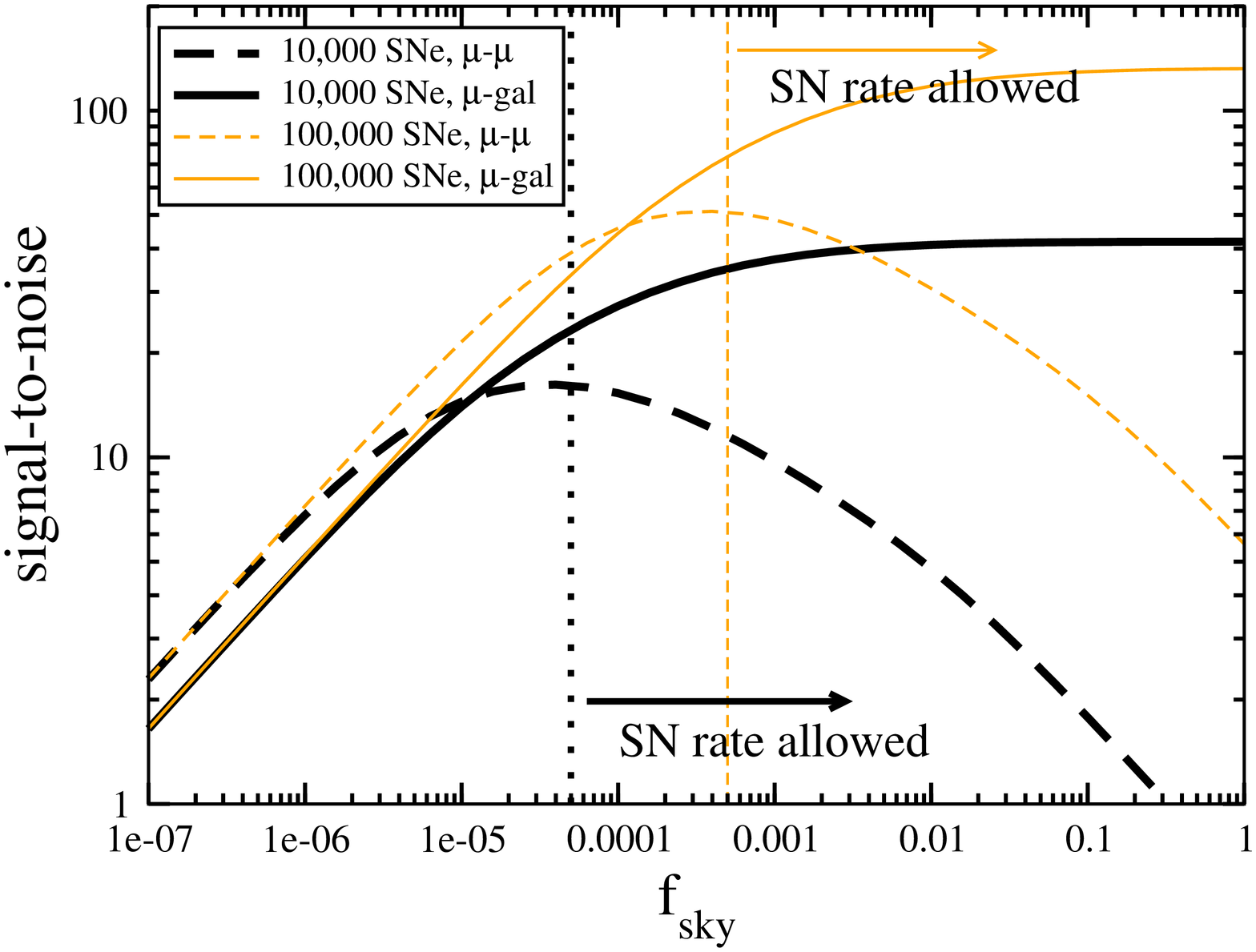,width=2.8in,angle=-90}}
\caption{{\it Left panel:} Signal-to-noise ratio for the detection of
the magnification power spectrum (dashed curves) and the magnification-galaxy cross
power spectrum (solid curves) as a function of the smallest scale (maximum
multipole) probed by the survey. We show cases of 3,000 SNe (cyan/light curves) and
10,000 SNe (black/dark curves)  collected over 20 sq.\ deg.\ ($\fsky\approx
0.0005$).  Note that the cross power spectrum can be detected with better
signal-to-noise than the magnification auto-correlation because the shot noise
in the foreground galaxy population is much smaller than that in the SNe
number density.  {\it Right panel:} Signal-to-noise ratio of the magnification
power spectrum (black/dark curves) and the galaxy-magnification cross-power
(orange/light curves) as a function of the fraction of the sky covered, $\fsky$, and
assuming 10,000 SNe (solid) and 100,000 SNe
(dashed). Vertical lines show the minimal
$\fsky$ in for a given number of SNe, which is given by the SN rate over the
survey area and assuming an observing time of five years.  }
\label{fig:sn}
\end{figure*}

In addition to a measurement of the projected angular power spectrum of cosmic
magnification, one can also cross-correlate the magnification with the foreground
galaxy distribution. The idea here is that the dark matter distribution that
causes the magnification pattern $\delta \mu(z, \hat{\bf n})$ is traced by
galaxies and, therefore, $\delta_{\mu}$ and the normalized galaxy overdensity,
$\delta_{\rm gal}$, are correlated.
The projected cross-correlation between the two fields is described by 
the angular power spectrum 
\begin{eqnarray}
C_\ell^{\mu-gal} &=& \int dr  \frac{W(r) n_{\rm gal}(r)}{d_A^2} P_{\rm dm-gal}
\left(k=\frac{\ell}{d_A},r\right) \, , \nonumber \\
\end{eqnarray}
where $n_{\rm gal}(r)$ is the normalized radial distribution of foreground
galaxies. Since $\delta_L=\delta_\mu$, the cross-correlation is
independent of the power-law slope of the source number counts, unlike in the case of
traditional galaxy-quasar cross-correlation measurements \citep{Scranton:05}.

To estimate how well these angular power spectra can be measured with upcoming
surveys, we compute the cumulative signal-to-noise ratio for detection
\begin{equation}
\left(\frac{{\rm S}}{{\rm N}}\right)^2 = \sum_\ell 
\left(\frac{C_\ell^i}{\Delta C_\ell^i}\right)^2 \, ,
\label{eq:sn}
\end{equation}
where the index $i$ references either the magnification power spectrum or the
magnification-galaxy cross power spectrum.  
The error in the magnification power spectrum is
given by
\begin{equation}
\Delta C_\ell^{\mu \mu} = \sqrt{\frac{2}{(2\ell+1)f_{\rm sky}\Delta \ell}} 
\left[ C_\ell^{\mu \mu} + \frac{\sigma^2_\mu}{N_{\rm SN}}\right] \, ,
\end{equation}
where $N_{\rm SN}$ is the surface density of SNe (number per steradian),
$\sigma_\mu$ is the uncertainty in the $\delta_\mu$ measurement from
each supernova, $f_{\rm sky}$ is the fraction of sky covered by the
survey,  and $\Delta\ell$ is the binning width in multipole space.
For the SN luminosity-galaxy count cross correlation, the error is
\begin{eqnarray}
\Delta C_\ell^{\mu-gal} &=& \sqrt{\frac{1}{(2l+1)f_{\rm sky}\Delta \ell}} \nonumber \\
&\times& \left[ \left(C_\ell^{\mu-gal}\right)^2 + 
\left(C_\ell^{\mu \mu}+\frac{\sigma^2_\mu}{N_{\rm SN}}\right)
\left(C_\ell^{gal-gal}+\frac{1}{N_{\rm gal}}\right) \right]^{1/2} \, ,
\end{eqnarray}
where $C_\ell^{gal-gal}$ is the angular clustering power spectrum of foreground
galaxies, and $N_{\rm gal}$ is their surface density.

For definitiveness we assume a magnification measurement error for each SN of
$\sigma_\mu=0.1$\footnote{The magnification error
is equal to the relative error in
measuring luminosity, and roughly equal to twice the relative
luminosity distance error.}; while smaller than errors
in current SN surveys, this is expected to be achievable in the near future.
For simplicity we consider $N_{\rm SN}$ SNe uniformly distributed in
$0.1\leq z \leq 1.7$, which roughly approximates the
distribution expected from SNAP \citep{snap}.

\section{Results and Discussion}
\label{sec:discussion}

Figure~\ref{fig:cl} shows the angular power spectra of cosmic magnification of
SNe, the cross power spectrum between the magnification anisotropies and the
foreground galaxy distribution, and the galaxy angular power spectrum. Here we
are particularly interested in the $\mu$--$\mu$ and $\mu$--gal spectra, for which
we also show error bars.  We have assumed the same foreground galaxy sample as in
\citet{Scranton:05}, with the redshift distribution of the form $dn/dz \sim
z^{1.3} \exp[-(z/0.26)^{2.17}]$ and a number density of usable galaxies of 3
arcmin$^{-2}$. To avoid overlap of SNe with the galaxy sample, in the case of
cross-correlation we only consider a SN subsample with redshifts greater than
0.7. In Figure~\ref{fig:cl} we also show the second order
correction to magnification corresponding to the $\kappa^2$ term in
Eq.~(\ref{eq:higher_order}); we calculate this following \citet{Menard:03}, but
using the halo model description for the density field bispectrum rather than a
numerical fit to simulations.

In Figure~\ref{fig:sn} we show the total signal-to-noise ratio for the
detection of the angular power spectrum of SN magnifications. The left panel
shows the S/N as a function of the smallest scale (maximum multipole) probed by
the survey. We show cases of 3,000 and 10,000 SNe observed over 20 sq.\ deg.\
(i.e.\ $\fsky \approx 0.0005$).  Note that, while the signal-to-noise ratio for
the magnification power spectrum is usually around 20 or below, the cross power
spectrum can be detected with considerably better significance due to the much
smaller shot noise in the foreground galaxy population.  Compare this to the
current state-of-the-art in mapping the cosmic magnification: the SDSS catalog
of quasars and galaxies has been used to detect the cosmic magnification at the
8$\sigma$ level, and this detection is not expected to improve significantly
given the already impressive statistics. In the future, SNe will provide the
best opportunity to extend and improve the mapping of the cosmic magnification.
The right panel of Figure~\ref{fig:sn} shows the signal-to-noise ratio, except
now as a function of the sky coverage of the survey, $\fsky$, as we hold the
total number of observed SNe fixed.  This allows us to optimize the
magnification measurement for a given amount of telescope time.  Very small
$\fsky$ leads to large cosmic variance, while large $\fsky$ decreases the
surface density of SNe (since both their number and the observation time are
held fixed) and therefore increases the shot noise. In addition, the upper
limit on the rate of SNe translates into a minimum $\fsky$. Measurements of the
actual rate of SNe \citep{Pain} combined with theoretical estimates
\citep{Oda:05} suggest that a year-long survey should find up to $\sim$ 10$^3$
SNe per square degree, and this limit, assuming a five-year survey, is shown as
a lower limit on $\fsky$ in the right panel of Figure~\ref{fig:sn}.

The surface density of galaxies, estimated to be around $10^9$ sr$^{-1}$ down
to 27th magnitude \citep{Smail:95}, is far larger than the surface density of
SNe (which is of order $10^6$ sr$^{-1}$ over a year-long
integration). Nevertheless, cosmological methods that use galaxies to probe the
formation of structure in the universe, chiefly through weak gravitational
lensing, are subject to systematic errors that range from theoretical
uncertainties to a variety of measurement systematics (see
\citealt{Hutereretal:05} and references therein).  Supernova measurements of
the magnification can be extremely valuable in helping control these
systematics and break certain degeneracies.

For example, in order to establish the masses of shear-selected galaxy
clusters, one can reconstruct the convergence $\kappa$ from the measured shear
maps, $\gamma_{1,2}(\hat{\bf n})$.  While well-known techniques exist for this
purpose (e.g.\ \citealt{Kaiser:93}), the reconstruction is insensitive to a
constant mass sheet (or spatially uniform convergence).  This mass-sheet
degeneracy (e.g.\ \citealt{Falco:85,Bradacetal04}) can be broken with direct
convergence measurements via magnification, and SNe are ideal candidates for
this purpose.  Updating the calculations in \citet{kolatt}, we
find that up to ~2\% of the SNe are magnified by foreground clusters at a
factor greater than 1.3 (a 3$\sigma$ or better detection).  A survey covering
$\sim$ 20 sq.\ deg.\ with $\sim 10^4$ SNe, combined with a shear
analysis, can provide mass (enclosed out to the impact radius of the background 
supernova) of $\sim$ 100 clusters to better than 10\%.

This approach can also test the consistency between shear measurements from
galaxy shapes and convergence from SNe luminosity anisotropies. One can
construct E- and B-modes of shear and, in the weak lensing limit,
$C_\ell^\kappa = C_\ell^E$ and $C_\ell^B=0$. Departures from these relations
are expected from both physical and theoretical systematic uncertainties.  For
example, intrinsic correlations between galaxies may produce additional but
unequal contribution to E and B-modes \citep{Heavens:00}. Moreover, there will
exist contributions from higher-order effects due to slight departures from the
weak lensing limit (see Eq.~(\ref{eq:higher_order}); \citealt{Menard:03}) or
higher order corrections to lensing that induce a rotational component via
coupling between two or more lenses \citep{CooHu02}, or due to the presence of
a gravitational wave background \citep{dodelson}. The power of this consistency
test is limited by the size of the higher order corrections.

To quantify the detectability of the difference between the shear and
convergence power spectra, we assume for a moment that this difference is given
by the second-order term $C_\ell^{\kappa^2-\kappa}$ plotted as the
dot-dashed line in Fig.~\ref{fig:cl}. We calculate the signal-to-noise in
measuring the quantity $C_\ell^\Delta=|C_\ell^\kappa-C_\ell^\gamma|$ following the
procedure similar to that in Eq.~(\ref{eq:sn}). Since future weak lensing shear
surveys will have much smaller shot noise ($\gamma_{\rm rms}^2/\bar{n}\sim
10^{-11}$) than the corresponding magnification power measurements ($\sim
10^{-9}$), the former source of noise can be ignored in the calculation.  Using
the SNe surface density of 10$^3$ deg$^{-2}$ yr$^{-1}$, we find that this
difference between the power spectra can be detected with a signal-to-noise
ratio of $\sim 10 \,(20 \; {\rm deg^{2}}/A)^{-1/2}$, where $A$ is the total
survey area.  Alternatively, if we assume that the fiducial difference between
the power spectra has a shot-noise power spectrum (i.e.  flat in $\ell$) with
$C_\ell^\Delta = \Delta$, the minimum detectable amplitude (with a signal-to-noise
ratio of unity) is $\Delta \approx 5 \times 10^{-7}\,(20 \; {\rm
deg^2}/A)^{1/2}$.  Consequently, corrections to the shear signal that are
due to intrinsic correlations may be detectable \citep{Jing:02}.

Magnification statistics from SNe also provide information on the cosmological
parameters. This can be estimated in similar fashion to the case of
conventional weak lensing of galaxies \citep{HuTeg:99,Huterer_2002}, since both
techniques probe the matter power spectrum and geometrical distance factors
(see Eq.~\ref{eq:cl}).  With the magnification power spectrum detected at a
signal-to-noise ratio of 10 (100), one linear combination of parameters,
typically with large weights in the $\Omega_m$ and $\sigma_8$ directions, can
be constrained to 10\% (1\%). While conventional weak lensing of galaxies can
provide a more accurate overall determination of cosmological parameters, the
strength of the proposed method is that it combines lensing shear and
magnification information in the same field, thereby providing a number of
cross checks on systematics.

\vspace{-0.5cm}
\acknowledgements D.H.\ is supported by an NSF Astronomy and Astrophysics
Postdoctoral Fellowship under Grant No.\ 0401066. D.E.H. acknowledges a Richard
P. Feynman Fellowship from Los Alamos National Laboratory.

\end{document}